# Optoelectronic properties of a photosystem I - carbon nanotube hybrid system

**Simone M Kaniber**[1,4], **Friedrich C Simmel**[2,4], **Alexander W Holleitner**[1,4] **and Itai Carmeli**[3,5]

[1] Walter Schottky Institut, Technische Universität München, Am Coulombwall 3, D-85748 Garching (Germany)

E-mail: holleitner@wsi.tum.de

[2] Physik Department, Technische Universität München, James Franck Str. 1, D-85748 Garching (Germany)

[3] Chemistry department and NIBN, Ben Gurion University, 84105 Be'er Sheva (Israel)

E-mail: itai@post.tau.ac.il

**Abstract.** The photoconductance properties of photosystem I (PSI) covalently bound to carbon nanotubes (CNTs) are measured. We demonstrate that the PSI forms active electronic junctions with the CNTs enabling control of the CNTs photoconductance by the PSI. In order to electrically contact the photoactive proteins, a cysteine mutant is generated at one end of the PSI by genetic engineering. The CNTs are covalently bound to this reactive group using carbodiimide chemistry. We detect an enhanced photoconductance signal of the hybrid material at photon wavelengths resonant to the absorption maxima of the PSI compared to non-resonant wavelengths. The measurements prove that it is feasible to integrate photosynthetic proteins into optoelectronic circuits at the nanoscale.
(Keywords: Carbon Nanotubes, Hybrid Nanosystem, Optoelectronics, Photosystem I (PSI)

## 1. Introduction

Photosynthetic reaction centers are the photochemically active complexes in photosynthetic systems found in plants, algae and photosynthetic bacteria [1]. The photosynthetic reaction centers have evolved approximately 3.5 billion years ago, and they serve as the ultimate source of energy in the biosphere. The photosynthetic process involves an efficient conversion of solar energy to stable chemical energy. During this process the reaction center photosystem I (PSI) acts as a nanosize photodiode composed of a protein chlorophyll complex that utilizes light to generate a photopotential of about 1 V with a quantum efficiency of 1 and an intrinsic energy conversion efficiency of 58% (47% of the total absorbed light) [2-7].

---

[4] LMU Munich, Geschwister-Scholl-Platz 1, D-80539 München (Germany)
[5] Department of Chemistry and Biochemistry, Tel-Aviv University, 69978 Tel-Aviv (Israel)

Recently, we have demonstrated the possibility to covalently bind the PSI reaction center directly to gold surfaces [4] as well as indirectly via a small linker molecule to GaAs surfaces [8] and to single- and multi-walled carbon nanotubes (CNTs) [3]. In ref. [3], we also demonstrated that self-assembled monolayers (SAMs) of PSI on top of an Au-electrode can be contacted by CNTs as well. We showedinitial evidence that the latter SAM-device shows an enhanced photoconductance signal due to the PSI [3]. In the present study, we investigate the optoelectronic properties of hybrids made of the PSI and single-walled carbon nanotubes in greater detail. In particular, the PSI–CNT hybrids are characterized by photoconductance measurements as a function of the excitation wavelength, the laser chopper frequency, and the excitation location. We detect a strongly enhanced photoconductance signal of the hybrid material at photon wavelengths resonant with the absorption maxima of the PSI compared to non-resonant wavelengths. We do not observe such an enhanced photoconductance signal in control samples consisting merely of purified CNTs. We consider charge and energy transfer processes from the PSI to the CNTs in order to explain the observed photoconductance response of the PSI-CNT hybrid system [9]. We further rule out photodesorption-effects on the surface of the CNTs [10], the effect of Schottky contacts between the CNTs and the metal contacts [11], electron-hole effects only within the CNTs [12-17], and bolometric effects [18] to dominate the photoconductance properties of the PSI-CNT hybrids. The results indicate that the integrated proteins are optoelectronically active, and that the electron transfer chain within the PSI can be electronically contacted by CNTs. This study demonstrates that the PSI has the potential to be integrated in nanoscale solar- or power-cells, and the electrical addressing via CNTs is a first step in this direction. Namely, the excellent optical and electronic properties of CNTs qualify them as promising building blocks of nanoscale optoelectronic devices [19-27], and the length of CNTs of up to several microns make them suitable as mesoscopic electrodes for the PSI as well as nanocrystals [28] and molecules [29,30].

## 2. Materials and Methods

*2.1. Material and Synthesis*

The PSI protein complex has a cylindrical shape with a diameter of about 15 nm and a height of 9 nm [1]. It is composed of polypeptide chains in which chlorophyll and carotenoids are imbedded (figure 1(a)). Following photo excitation, an electron is transferred along an electron transfer pathway within the protein complex (depicted in red, figure 1(a)) from the lumen to the stroma side of the PSI. The final acceptors are located about 6 nm away from the initially oxidized reaction-center chlorophyll P700. Intriguing for optoelectronic applications, the photoexcited state of the PSI provides a surface photovoltage of up to 1 V [2-4]. In the studied PSI, amino acids in the extra membrane loops of the PSI facing the lumen side of the bacterial membrane (oxidizing side) are mutated to cysteines (Cys); enabling the formation of covalent bonds with a solid state surface [4,8] or CNTs [3]. The Cys located at extra membranal loops of the protein do not have steric hindrance, when placed on a solid surface. The mutations D235C/Y634C are selected near the special chlorophyll pair P700, indicated by the black arrow in figure 1(a), to allow close proximity between the reaction center and a solid state surface. Our assembly approach facilitates efficient electronic junctions and avoids disturbance in the function of the reaction center. The covalent attachment of the PSI through the Cys further ensures the structural stability of the assembled, oriented PSI–CNT hybrids.

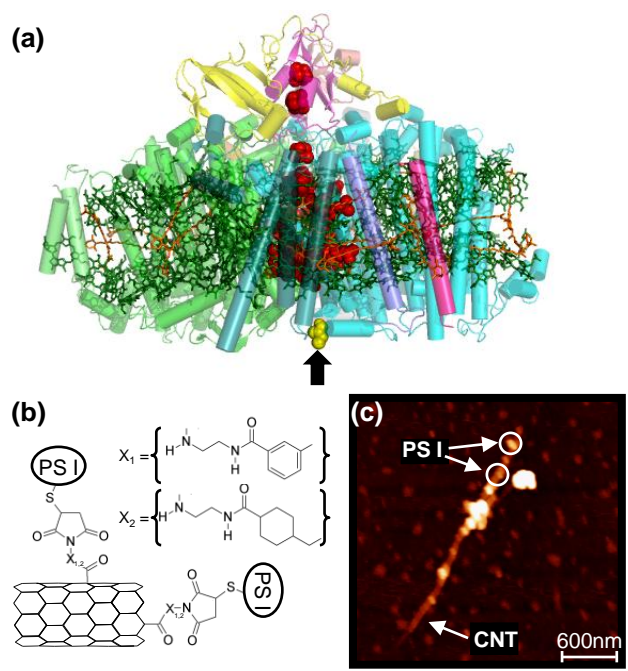

**Figure 1.** a) Molecular structure image of the photosystem I (PSI)**.** The PSI covalently binds to maleimide modified carbon nanotubes through a Cys mutation on the polypeptide backbone (black arrow and space fill model, yellow) which is close to the chromophores that mediate the electron transfer (represented by the space fill model (red)). b) Scheme of a nanoscale hybrid composed of carbon nanotubes (CNTs) and the (PSI). The sulfur containing cystein group of a mutant PSI binds covalently to a carboxylated CNT via a maleimide linker [3]. c) Atomic force micrograph (AFM) of PSI–CNT hybrids. The image shows a large number of PSI (circles) with a diameter of about 9-20 nm bound to the tips and the side-walls of a CNT bundle [3].

The functionalization of CNTs by chemical modifications holds interesting prospects in various fields such as nucleic acid sensing [31] and the fabrication of hybrid bioorganic nanosystems [32-37]. In particular, it has been demonstrated how to covalently bind DNA [33] and nanocrystals [34, 35, 38-40] to CNTs, as well as CNTs to the end of an AFM tip [41]. In a previous study [3] we have demonstrated the possibility to covalently bind PSI to CNTs by a four step chemical route utilizing sulfo-SMCC (sulfosuccinimidyl 4-[N-maleimidomethyl] - cyclohexane-1-carboxylate). Here, we utilize sulfo-MBS (m-Maleimidobenzoyl-N-hydroxysulfosuccinimide ester) instead of sulfo-SMCC, which results in a linker molecule that has an aromatic hydrocarbon instead of a cyclohexane (figure 1(b)). We found that the difference between the two linker molecules does not result in modifications of the optoelectronic properties of the hybrid systems. Most importantly, the covalent binding of the Cys mutations located at the lumen side of the PSI to the CNTs give rise to a unique topology of the PSI–CNT hybrids for optoelectronic applications. In this configuration, the oxidizing side of the PSI is oriented towards the CNTs.

*2.2 Characterization*

To evaluate the degree of conjugation between the modified CNTs and PSI, a drop of the solution is placed onto a silicon surface and incubated for two hours. The samples are washed briefly with deionized water and dried under nitrogen. Figure 1(c) shows an atomic force micrograph (AFM) of a PSI–CNT hybrid on a silicon surface. The image exhibits a large number of spherical particles (circles in figure 1(c)) attached to the surface of the CNTs. The height analysis of the AFM images indicates a height of about 9 - 20 nm for the spherical particles (data not shown), in agreement with the actual diameter of the PSI which suggests that the spherical particles are the PSI proteins [3]. The diameter of the CNTs is in the range of a few up to 6 nm. Single-walled CNTs typically have a diameter of about 1 - 2 nm [19], while a height of 6 nm suggests that bundles of CNTs build the back-bone of the PSI–CNT hybrids [42].

*2.3 Photoconductance measurements*

We electronically contact ensembles of PSI–CNT hybrids and purified CNTs by depositing a drop of an aqueous solution, which contains PSI–CNT hybrids or purified CNTs, onto an insulating $SiO_2$ substrate with opto-lithographically predefined gold contacts on top. The metal contacts are electronically bridged by the hybrids after drying. Because the area of the dried hybrid nanosystems is smaller than the metal contacts, the contacts act as source-drain electrodes for the PSI–CNT hybrids in a two-terminal configuration (sketch in figure 2(a)).

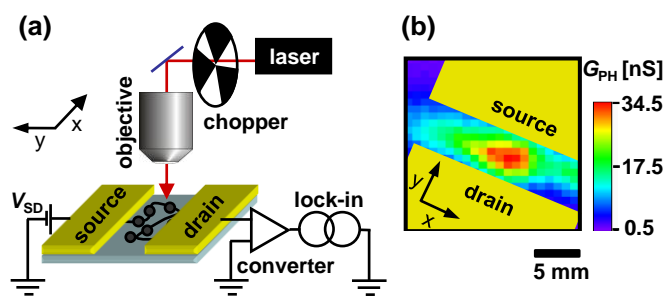

**Figure 2.** a) Experimental circuit for measuring the photoconductance of PSI–CNT hybrids as a function of the spatial coordinates *x* and *y* (see section 2.3 for details). b) Spatially resolved

photoconductance signal at room temperature of an ensemble of PSI–CNT hybrids, which electronically bridge metal source-drain contacts with a distance of about 5 μm ($\lambda$ = 420 nm).

The photoconductance $G_{PH}(\lambda, f_{CHOP})$ of the hybrids is measured as a function of the photon wavelength $\lambda$, the chopper frequency $f_{CHOP}$, and the location of the laser excitation with respect to the source-drain contacts (see sketch of the experimental circuit in figure 2(a)). To this end, a bias voltage $V_{SD}$ is applied across the source-drain electrodes, while the light of a titanium:sapphire laser is focused through an objective of a microscope onto the hybrid nanostructures. The current across the sample is fed through a current-voltage converter. A lock-in amplifier provides a dc-signal of the current difference $\Delta I = I_{ON} - I_{OFF}$ across the sample for the laser being "on" or "off". For a modulation frequency below 2 kHz, the necessary reference trigger is given by the chopper frequency $f_{CHOP}$ of the chopper wheel (figure 2(a)). For the photoconductance measurements at 76 MHz, we utilize the repetition frequency of the mode-locked titanium:sapphire laser. In both cases, the resulting $\Delta I$ can be translated into a photoconductance $G_{PH} = \Delta I / V_{SD} = G_{PH}(\lambda, f_{CHOP})$ of the electronically contacted nanosystem because the samples measured show an ohmic behavior at small $V_{SD}$. The zero-potential of the current-voltage converter ensures that the source-drain voltage $V_{SD}$ only drops across the sample. The conductance $G$ is defined by $G = I / V_{SD}$. The location of the objective and thus the laser excitation with respect to the source-drain contacts can be controlled by piezo positioners at a step size of about 50 nm. We utilize the current-voltage converter (lock-in) Ithaco 1211 (eg&g 7260) at frequencies below 2 kHz and a FEMTO HCA-100M_50K-C (Stanford Research SR 844) at a frequency of 76 MHz.

*2.4 Absorbance Measurements*

The absorbance of the PSI is measured spectrally resolved with an UV-VIS spectrophotometer Jasco V-550. The absorption intensities for two different samples, one containing PSI in a tricine buffer solution (*I*) and one containing pristine tricine buffer solution (PH 7.5 - 8) as a reference ($I_0$), are detected and compared to each other. The absorbance A is given by the formula $A = - \log (I / I_0)$.

*3. Experimental Results*

There are several processes which can alter the photoconductance of the PSI–CNT hybrids, such as photodesorption-effects on the surface of the CNTs [10], the effect of Schottky contacts between the CNTs and the gold contacts [11], electron-hole effects within the CNTs [12-17], bolometric effects [18], and charge as well as energy transfer processes within the PSI–CNT hybrids [9]. First, all of our samples are measured in vacuum ($p < 1 \times 10^{-3}$ mbar). Hereby, we can rule out photodesorption effects, where the laser excitation would induce oxygen desorption of the dopant oxygen from the side-walls of the CNTs [10]. To rule out the effect of a Schottky barrier between the CNTs and the gold contacts as the dominating optoelectronic effect, we measure the photoconductance of ensembles of PSI–CNT hybrids, while we spatially scan the excitation spot in the directions *x* and *y* with respect to the source-drain contacts (figure 2(a) and (b)). We find that the photoconductance exhibits a maximum in the center of the source and drain contacts at a distance of about 2 μm away from the metal contacts (figure 2(b)). Since Schottky barriers only emerge in close vicinity to the metal contacts, the data of figure 2(b) is evidence that Schottky barriers do not dominate the photoconductance in the present PSI–CNT hybrids. Comparing the spatial dependence of several PSI-CNT hybrid samples we determine that in all cases, the signal is observed at positions where ensembles of dried PSI-CNT hybrids are located.

In figure 3(a) and (b), the normalized photoconductance $G_{PH}$ of an ensemble of purified CNTs without PSI is depicted as a function of the excitation wavelength [43]. We find a maximum of $G_{PH}$ at about $\lambda = 822 \pm 11$ nm (red lines are Gaussian fits to the experimental data). Following recent reports [12-17], we interpret the finding in figure 3(a) and (b) such that electron-hole dynamics in the semiconducting CNTs of the ensemble give rise to the observed photoconductance. A wavelength of $\lambda = 822 \pm 11$ nm translates into an intersubband transition $E_{22} = 1.51 \pm 0.16$ eV of a semiconducting single-walled CNT with a diameter of $d = 1.11 \pm 0.06$ nm (and $E_{11} = 0.77 \pm 0.07$ eV) [44]. We would like to note that the derived diameter of the CNTs is consistent with Raman data on the same set of CNTs (data not shown). Because the absorption cut-off of the PSI is around 700 nm (~ 1.77 eV), the absorption corresponding to the $E_{22}$ transition in the CNTs are detected in the photoconductance signal of the PSI–CNT hybrids at a photon energy smaller than 1.77 eV as well, but they are not assigned to any contributions from the PSI. Indeed, for $\lambda > 780$ nm the photoconductance signal of the PSI–CNT hybrids rises for a longer wavelength (black triangle in figure 3(c)). Intriguingly, for $\lambda < 780$ nm we detect a photoconductance signal of the PSI–CNT hybrids, which mimics the absorbance spectrum of the PSI to first order (red line in figure 3(c)). The absorbance spectrum of the PSI is measured with the proteins being suspended in a buffer solution (see section 2.4). Typical for PSI, the chlorophyll absorbance maxima are at ~ 440 nm and 680 nm, and the carotenoids contribute to a shoulder at 500 nm.

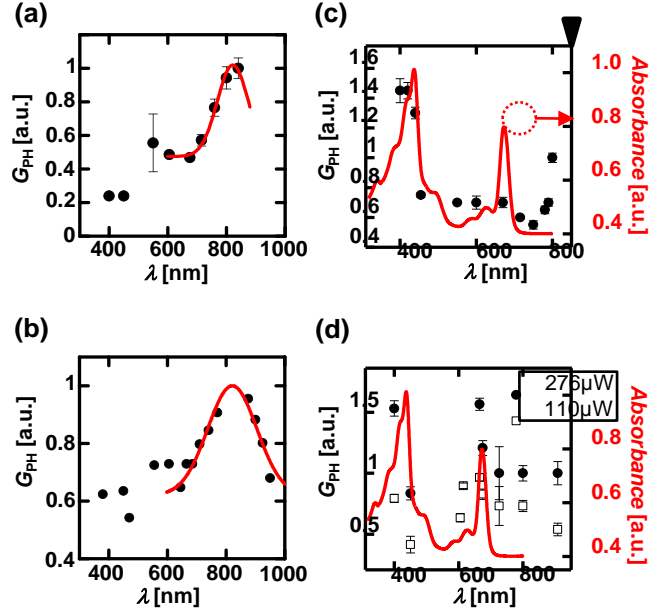

**Figure 3.** a) Normalized photoconductance of purified CNTs as a function of the photon wavelength $\lambda$ at $f_{CHOP}$ = 1.1 kHz and $I_{LASER}$ = 32 kWcm$^{-2}$ at room temperature [43]. b) Normalized photoconductance spectrum of pristine CNTs measured at 76 MHz and $I_{LASER}$ = 28 kWcm$^{-2}$. The red lines in a) and b) are Gaussian fits to the data with a center wavelength of 822 nm and 821 nm. c) Normalized photoconductance of the PSI-CNT hybrids at $f_{CHOP}$ = 1.1 kHz and $I_{LASER}$ = 6.4 kWcm$^{-2}$. d) Normalized photoconductance data on the same sample as in c) at 76 MHz and $I_{LASER}$ = 35 kWcm$^{-2}$ and 14 kWcm$^{-2}$ (110 μW and 276 μW). Red Lines in c) and d): Normalized absorbance of PSI in a buffer solution. See text for details.

As can be seen from figure 3(c), both the contribution of the chlorophyll and the carotenoids are observed in the photoconductance signal of the PSI–CNT hybrids for $f_{CHOP}$ in the kHz regime. The maximum at 680 nm is slightly suppressed as compared to the absorbance spectrum (dashed circle). At a trigger frequency of 76 MHz for the lock-in amplifier (see section 2.3), the maximum at 680 nm can also be resolved (figure 3(d)), which we do not detect for purified CNTs at kHz (figure 3(a)) and 76 MHz (figure 3(b)). We interpret the finding such that the optoelectronic response of the PSI-CNT hybrids happens on a fast time scale at 680 nm. As described in section 2.3, $G_{PH}$ measures the difference of the conductance between the on and off state of the laser, and thereby it is sensitive to processes occurring at the chopper frequency.

Figure 4(a) shows a typical frequency dependence of the photoconductance of the PSI–CNT hybrids in the kHz regime. For all $\lambda$ and within the experimental error, we find no frequency-dependence of the photoconductance for both the PSI–CNT hybrids and the purified CNT samples at a chopper frequency up to 2 kHz. Thus, we can exclude photoconductance dynamics within the PSI–CNT hybrids in the millisecond time-scale (1/ (2 kHz) = 0.5 ms). In turn, bolometric effects due to a change of the lattice temperature within the CNTs can be ruled out to dominate the photoconductance in the CNTs because such bolometric effects occur on the millisecond time-scale [18].

Figure 4(b) demonstrates that there are charge carrier dynamics within the PSI–CNT hybrids, which occur on a time-scale of seconds and longer. The upper graph of figure 4(b) depicts the photoconductance $G_{PH}$ of the PSI–CNT hybrids as a function of the laboratory time. As noted above, whenever the laser is switched on, a constant photoconductance signal is measured (shaded area in figure 4(b)), while if it is switched off there is no signal $G_{PH}$ detected. As can be viewed in the lower graph of figure 4(b), after the laser excitation a stepwise change in $G$ with a time-scale of several seconds is observed (see exponential fit in the lower graph of figure 4(b)). The conductance $G$ reaches the value of the initial state only after the PSI–CNT hybrids having been in darkness for several hours.

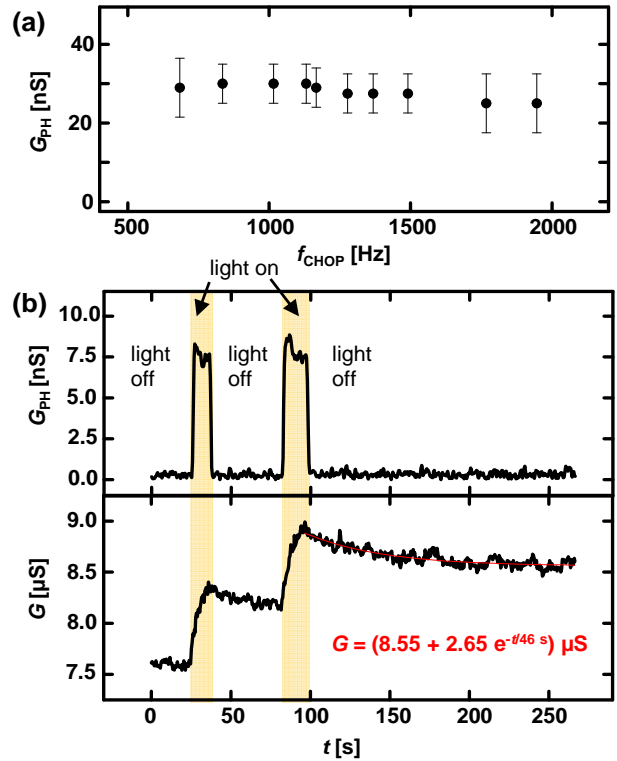

**Figure 4.** a) Typical frequency dependence of the photoconductance of the PSI–CNT hybrids in the kHz regime at $I_{LASER} = 17$ kWcm$^{-2}$ at room temperature. b) Upper graph: photoconductance $G_{PH}$ of the PSI–CNT hybrids as a function of the laboratory time $t$. Lower graph: conductance $G$ measured at the same time as the upper graph. Red line is an exponential fit to the data with a typical time constant of several seconds.

## 4. Discussion

The data in figure 3(c) and (d) demonstrate that the PSI-CNT hybrid material exhibit a strongly enhanced photoconductance signal at photon wavelengths close to the absorption maxima of the PSI compared to non-resonant wavelengths. In addition, the control samples of purified CNTs do not show such an enhanced photoconductance signal (figure 3(a) and (b)). The data in figure 2(b) further verifies that the conductance change $G$ in the CNTs is induced by optically excited PSI proteins, which are bound to the CNTs. Therefore, the results demonstrate that an electronically active junction is formed between the PSI and the CNTs, which influences the photoconductance of the CNTs. The PSI-CNT hybrids exhibit a unique topology in which the oxidizing side of the PSI is oriented toward the CNTs. Hereby, the excitation of the PSI by light induces a vectorial charge separation perpendicular to the CNTs surface. The electron is transferred along the electron transfer chain of the reaction center (figure 1(a)), leaving a hole at the oxidizing side of the PSI which is coupled to the CNT surface. It is therefore likely that the change in the photoconductance of the PSI-CNT hybrids is caused by the optically excited hole which tunnels into the CNTs. Such a charge transfer gives rise to a change of the charge density $n$ and in turn, a change of the conductance $G$ of the CNTs ($G \propto \sigma = nq_e\mu$, with $\sigma$ the conductivity of the CNTs, $n$ the charge carrier density of the CNTs, $\mu$ the carrier mobility of the CNTs, and $q_e$ the electron charge). In this scenario, the optically excited electron stays at the final acceptor within the PSI until it eventually recombines with the spatially separated hole. Most importantly, this interpretation explains the long time-scale of the optically induced conductance change in the lower graph of figure 4(b) in a way that the time-scale reflects the recombination of the spatially separated charge carriers. A different interpretation, in which the hole resides inside the PSI with no charge transfer to the CNT, is unlikely, since such an overall optoelectronic process would have a time constant which is comparable to the recombination time of electron-hole excitations in pristine PSI of about milliseconds [1]. This is not the case for the hybrid system as verified by the data presented in figure 4(a). Therefore, the presented results indicate the formation of an electronically active junction which mediates a vectorial light induced charge transfer between the bound PSI and the CNTs. Further experimental and theoretical work is underway to determine the specific factors that govern charge transfer dynamics in the PSI-CNT hybrids.

## 5. Conclusion

In conclusion, genetically mutated PSI containing cysteine amino acids on specific sites of the PSI are utilized in order to construct hybrid protein junction with a unique topology; namely the PSI is covalently bound to the CNTs with the oxidizing side of the PSI oriented towards the CNTs. The conjugated system demonstrates intriguing optoelectronic properties which emerge from the formation of an active electronic junction between the PS I and the CNTs. In particular, we observe a photoconductive gain effect which follows the absorbance spectrum of the PSI. The study demonstrates that the PSI-CNT hybrids act as nanosized optical components which have the potential to serve as optical switches, nanoscale solar-cells and sensors for light.

**Acknowledgements**

We thank L. Frolov, S. Richter, C. Carmeli, and J.P. Kotthaus for fruitful discussions and support. We gratefully acknowledge financial support by the DFG (Ho 3324/2), SFB 486, the Center for NanoScience (CeNS), the LMUexcellence program and the German excellence initiative via the "Nanosystems Initiative Munich" (NIM).